\documentclass{revtex4}
\usepackage{graphicx}
\usepackage{fancyhdr}
\usepackage{amsmath}
\usepackage{color}

\begin{document}

\title{Scenarios with Composite Higgs Bosons}

%

\author{Michio Hashimoto}
\affiliation{Chubu University, 1200 Matsumoto-cho, Kasugai-shi, 
 Aichi, 487-8501, JAPAN}

\begin{abstract}
Typical models with composite Higgs bosons are briefly reviewed.
We also introduce the isospin symmetric Higgs model recently proposed 
in Ref.~\cite{Hashimoto:2012qe}. 
\end{abstract}

\maketitle

\thispagestyle{fancy}


\section{Introduction}

Recently, the ATLAS and the CMS Collaborations 
at the Large Hadron Collider (LHC) discovered 
a new boson $h$ in the mass range 125--126~GeV~\cite{LHC}.
In addition, the Standard Model (SM) Higgs boson has been excluded 
at 95\% C.L. in the mass range 110--149~GeV, except for the narrow region 
122.8--127.8~GeV~\cite{LHC-moriond}.
The mass range from 127 to 600~GeV was previously 
excluded~\cite{Chatrchyan:2012tx}. 
It is also noticeable that the mass $m_h=125$--$126$~GeV perfectly 
agrees with the LEP precision measurements~\cite{LEP}.
On the other hand, the contact interactions in the processes of
$pp \to \mbox{jet}$ and $pp \to \ell^+ \ell^-$ are severely
constrained, i.e., the compositeness scale $\Lambda$ should be 
larger than, say, 10~TeV~\cite{compositeness}.
Against this situation, is there still a room for some strong dynamics 
responsible for the electroweak symmetry breaking (EWSB)?

We give an overview of typical models with composite Higgs bosons.
It might give some hint for the origin of the EWSB.
We will also introduce the isospin symmetric Higgs model, 
which is recently proposed in Ref.~\cite{Hashimoto:2012qe}, 
as an example of the dynamical EWSB scenario. 



\section{dynamical EWSB}

The earliest idea of the dynamical EWSB is Technicolor (TC)~\cite{TC}:
The chiral condensate of (techni-) fermions is dynamically generated 
by the technicolor gauge interaction and 
it breaks the electroweak gauge symmetry, as in low-energy QCD.
The would-be Nambu-Goldstone (NG) bosons 
are eaten by the weak gauge bosons.
Then $W$ and $Z$ acquire their masses proportional to the technipion
decay constant ($\sim$ 100~GeV), which is analogous to 
the pion decay constant $f_\pi \simeq 93$~MeV in QCD.
In order to produce the masses of the SM fermions,
the extended technicolor (ETC) has been 
proposed~\cite{Dimopoulos:1979es,Eichten:1979ah}.
Although it was beautiful, this old-fashioned TC has been already 
excluded by several reasons~\cite{miransky-textbook,Hill:2002ap}.
For example, the constraint of the $S$-parameter~\cite{Peskin-Takeuchi} 
rules out this QCD-like TC with many weak doublets of 
the technifermions~\cite{pdg}.

A modern version of TC is the walking TC 
where the gauge coupling of the TC gauge group runs very slowly, 
or ``walks''~\cite{miransky-textbook,Hill:2002ap}.
The walking TC resolves the difficulties of the old-fashioned TC.
The estimate of the $S$-parameter expected from QCD
is not applicable to walking TC. 
Even in walking TC, however, it is difficult to generate 
the observed mass of the top quark from the ETC interactions 
without producing unacceptably large isospin breaking.
Also, the walking TC usually predicts a heavy composite Higgs boson.
Noticing that the $S$-parameter constraint requires 
a heavy technirho bound state $\rho_T$,
it is quite nontrivial to get a light composite Higgs, $m_h \sim 100$~GeV,
and simultaneously to obtain the heavy $\rho_T$, say, 
$M_{\rho_T} \gtrsim $ few TeV.

To generate the mass of the top quark, the topcolor dynamics is 
useful~\cite{Hill:2002ap}.
In the topcolor scenario, we assume that the new topcolor interaction
strongly couples to the third generation of quarks and then
the top quark condensate mainly yields the top quark mass.
In the simplest four dimensional model that the top condensate
is responsible both for the EWSB and $m_t$, and that
there appears only one (composite) Higgs doublet, 
too large top quark mass is predicted, however.
In a model with extra dimensions, 
this difficulty is relaxed~\cite{Hashimoto:2000uk}.
Another approach to avoid too large top mass is to assume
that the top quark condensate is responsible only for $m_t$
and the EWSB takes place by some other mechanism.
In topcolor assisted technicolor (TC2), the TC interaction
causes the EWSB~\cite{Hill:1994hp}.
Because the chiral symmetry is extended in TC2,
there appears the extra NG bosons, so-called top-pions.
The masses of the top-pions are generally light and thus
they are severely constrained.
In a model in Ref.~\cite{Hashimoto:2009xi},
we employed a subcritical dynamics (although nearcritical, i.e., strong) 
for the topcolor interaction, so that 
the mass of the scalar bound state of the top and the anti-top quarks
is naturally heavy.
This mechanism is used in the isospin symmetric (IS) Higgs model
recently proposed in Ref.~\cite{Hashimoto:2012qe}. 

We introduce the IS Higgs model in the next section.

\section{IS Higgs Model and its predictions}

The ATLAS and CMS experiments did not only announced the mass
of the new discovered boson $h$, 
but they also reported the nature of $h$:
While the decay channels of $h \to ZZ^{*}$ and $h \to WW^{*}$
are fairly consistent with the SM, the diphoton branching ratio 
Br$(h \to \gamma \gamma)$ is about 1.6 times larger than 
the SM value~\cite{LHC}.
In the latest results, the ATLAS collaboration confirmed 
the similar enhancement in the diphoton channel~\cite{ATLAS-moriond}.
On the other hand, the CMS group changed their previous results 
of the signal strength from $\sigma/\sigma_{\rm SM} = 1.6 \pm 0.4$ to
$\sigma/\sigma_{\rm SM} = 0.78^{+0.28}_{-0.26}$ for the mass-fit-MVA analysis and
$\sigma/\sigma_{\rm SM} = 1.11^{+0.32}_{-0.31}$
for the cut-based analysis~\cite{LHC-moriond}.
The situation thus becomes unclear at present.
In any case, the deviation from the SM in the diphoton channel, 
if established, would be an indication of a new physics beyond the SM.

Let us introduce the IS Higgs boson model.

The main characteristics of the IS Higgs boson model are 
as follows~\cite{Hashimoto:2012qe,Hashimoto:2009xi}:
a) It is assumed that the dynamics primarily responsible for 
the EWSB leads to the mass 
spectrum of quarks with no (or weak) isospin violation.
{\it Moreover, it is assumed that the values of these masses are 
of the order of the observed masses of the down-type quarks.} 
b) The second (central) assumption is introducing 
the horizontal interactions for the quarks in the three families. 
As a first step, a {\it subcritical} (although nearcritical, i.e., strong) 
diagonal horizontal interactions for the top quark is utilized 
which lead to the observed ratio $\frac{m_t}{m_b} \simeq 41.5$~\cite{pdg}.
The second step is introducing {\it equal} strength horizontal 
flavor-changing-neutral (FCN) interactions between 
the $t$ and $c$ quarks and the $b$ and $s$ ones. 
As was shown in Ref.~\cite {Hashimoto:2009xi}, 
these interactions naturally provide the observed ratio 
$m_c/m_s \simeq 13.4$ in the second family~\cite{pdg}. 
As to the mild isospin violation in the first family, 
it was studied together with the effects of the family mixing, reflected 
in the Cabibbo-Kobayashi-Maskawa (CKM) matrix~\cite{Hashimoto:2009xi}.

In this scenario, 
the main source of the isospin violation is only 
the strong top quark interactions. 
However, because these interactions are subcritical, 
the top quark plays a minor role in EWSB.
This distinguishes the IS Higgs scenario from the top quark condensate 
model~\cite{Hill:2002ap}.

One of the signatures of this scenario is the appearance of a
composite top-Higgs boson $h_t$
composed of the quarks and
antiquarks of the third family~\cite{Hashimoto:2009xi}.
Note that unlike TC2~\cite{Hill:1994hp},
this class of models utilizes subcritical dynamics
for the top quark, so that the top-Higgs $h_t$ is heavy in general.
Here we also emphasize that while the top-Higgs boson $h_t$ has 
a large top-Yukawa coupling,
the IS Higgs boson $h$ does not, $y_t \simeq y_b \sim 10^{-2}$. 
On the other hand, the $hWW^{*}$ and $hZZ^{*}$ coupling constants are
close to those in the SM.
Also, the mixing between $h$ and much heavier $h_t$ should be small. 

We now describe the decay processes of the IS Higgs $h$.

It is well known that the $W$-loop contribution to $H \to \gamma\gamma$
is dominant in the SM, while the top-loop effect is destructive
against the $W$-loop. 
In the IS Higgs model, however, the Yukawa coupling between the top and 
the IS Higgs $h$ is as small as the bottom Yukawa coupling,
so that the top-loop contribution is strongly suppressed.
The partial decay width of $h \to \gamma\gamma$ is thus enhanced 
without changing essentially $h \to ZZ^{*}$ and $h \to WW^{*}$.
A rough estimate taking the isospin symmetric 
top and bottom Yukawa couplings $y_t \simeq y_b \approx 10^{-2}$
is as follows:
\begin{equation}
\label{gamma}  
  \frac{\Gamma^{\rm IS} (h \to \gamma \gamma)}
       {\Gamma^{\rm SM}(H \to \gamma \gamma)} \simeq 1.56 , \quad
  \frac{\Gamma^{\rm IS} (h \to WW^{*})}
       {\Gamma^{\rm SM}(H \to WW^{*})} = 
  \frac{\Gamma^{\rm IS} (h \to ZZ^{*})}
       {\Gamma^{\rm SM}(H \to ZZ^{*})} = 
       \left(\frac{v_h}{v}\right)^2 \simeq 0.96 . 
\end{equation}
Here using the Pagels-Stokar formula \cite{PS}, we estimated 
the vacuum expectation value (VEV) of the top-Higgs $h_t$ 
as $v_t = 50$ GeV, and
the VEV $v_h$ of the IS Higgs $h$ is given by the relation $v^2 = v_h^2 + v_t^2$
with $v = 246$ GeV. Note that
the values of the ratios in Eq. (\ref{gamma}) are not very sensitive to 
the value of $v_t$, e.g., for $v_t=40$--$100$ GeV,
the suppression factor in the pair decay modes to $WW^*$ and $ZZ^*$
is $0.97$--$0.84$ and 
the VEV $v_h$ of the IS Higgs $h$ is given by the relation $v^2 = v_h^2 + v_t^2$
with $v = 246$ GeV. Note that
the values of the ratios in Eq. (\ref{gamma}) are not very sensitive to 
the value of $v_t$, e.g., for $v_t=40$--$100$ GeV,
the suppression factor in the pair decay modes to $WW^*$ and $ZZ^*$
is $0.97$--$0.84$ and 
the enhancement factor in the diphoton channel is $1.58$--$1.37$. 
For the decay mode of $h \to Z\gamma$, this model yields 
\begin{equation}
\label{Zgamma}
  \frac{\Gamma^{\rm IS} (h \to Z \gamma)}
       {\Gamma^{\rm SM}(H \to Z \gamma)} \simeq 1.07 \, .
\end{equation}

The values in Eq. (\ref{gamma}) agree well with the data in 
the ATLAS and CMS experiments.
However, obviously, the main production mechanism of the Higgs boson,
the gluon fusion process $gg \to h$, is now in trouble.
The presence of new chargeless colored particles, which considered by
several authors \cite{adj-S}
can help to resolve this problem.
We pursue this possibility in the next section.

\section{Benchmark Model with colored scalar}
\label{4}

As a benchmark model, 
we may introduce a real scalar field $S$ in the adjoint
representation of the color $SU(3)_c$:
\begin{eqnarray}
  {\cal L} \supset 
  {\cal L}_S = \frac{1}{2} (D_\mu S)^2 - \frac{1}{2} m_{0,S}^2 S^2 
  - \frac{\lambda_S}{4} S^4
  - \frac{\lambda_{h S}}{2} S^2 \Phi_h^\dagger \Phi_h,
  \label{Lag-S}
\end{eqnarray}
where $\Phi_h$ represents the IS Higgs doublet.
The effective Lagrangian ${\cal L}$ also contains
the IS Higgs quartic couplings $\lambda_h$, 
${\cal L} \supset - \lambda_h |\Phi_h|^4$.
The IS Higgs mass is $m_h = \sqrt{2\lambda_h} v_h$, and 
we will take it to be equal to $125$ GeV. 
The mass-squared term for the scalar $S$ is 
$  M_S^2 = m_{0,S}^2 + \frac{\lambda_{hS}}{2} v_h^2$,
and should be positive in order to avoid the color symmetry breaking.
Typically, $M_S \sim 200$~GeV. 

Taking into account the $S$ contribution to $gg \to h$,
we find appropriate values of the Higgs-portal coupling,
\begin{equation}
  \lambda_{hS} \simeq 2.5 \mbox{--} 2.7 \times \frac{M_S^2}{v v_h} \,.
\end{equation}
As a typical value, we may take $\lambda_{hS} = 1.8$ for 
$M_S= 200$~GeV and $v_t=50$~GeV.

A comment concerning the IS Higgs  quartic coupling $\lambda_h$ is in order.
In the SM, the Higgs mass 125~GeV suggests that the theory is 
perturbative up to an extremely high energy scale~\cite{RGE-lam}.
On the contrary, in the present model, when we take a large Higgs-portal 
coupling $\lambda_{hS}$ that reproduces $gg \to h$ correctly, 
the quartic coupling $\lambda_h$ will grow because
the $\beta$-function for $\lambda_h$ contains the $\lambda_{hS}^2$ term.
Also, there is no large negative contribution to 
the $\beta$-function for $\lambda_h$ from 
the top-Yukawa coupling $y_t \sim 10^{-2}$.

\begin{figure}[t]
  \begin{center}
  \includegraphics[width=7.5cm]{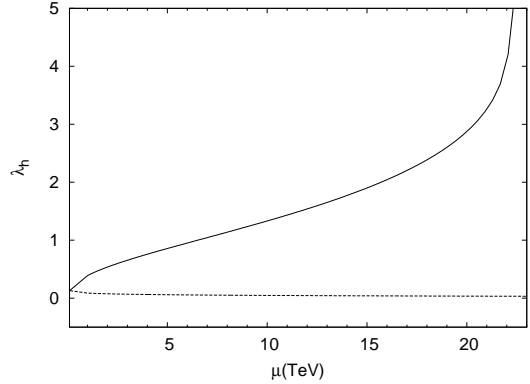}
   \end{center}
   \caption{The running behavior of the IS Higgs quartic coupling $\lambda_h$.
   The solid and dashed lines correspond to $\lambda_h$
   and the SM Higgs quartic coupling, respectively.
   We fixed the IS Higgs mass $m_h = \sqrt{2\lambda_h} v_h = 125$~GeV and
   took $\lambda_{hS}=1.8$ and $\lambda_S=1.5$.
   Unlike the SM, the IS Higgs quartic coupling grows up due to
   a large Higgs-portal coupling $\lambda_{hS}$  and a small
   top-Yukawa coupling $y_t$.
   \label{lam}}
\end{figure}

One can demonstrate such a behavior more explicitly by using the 
renormalization group equations.
In Fig.~\ref{lam}, the running of the coupling $\lambda_h$ is shown.
Taking a large Higgs-portal coupling $\lambda_{hS}=1.8$ and 
the $S^4$-coupling $\lambda_S=1.5$, 
it turns out that the coupling $\lambda_h$ rapidly grows.
The blowup scale strongly depends on the initial values of $\lambda_{hS}$ 
and $\lambda_S$.
A detailed analysis will be performed elsewhere.

Last but not least, we would like to mention that 
other realizations of the enhancement of the $h$ production 
are also possible.

\section{Conclusion}
\label{5}

We gave the overview of the typical scenarios with
the dynamical EWSB and also introduced the IS Higgs boson model.

In particular, the IS Higgs model can explain 
the enhanced Higgs diphoton decay rate observed at the LHC, 
and also makes several predictions. The most important of them is that
the value of the top-Yukawa coupling $h$-$t$-$\bar{t}$ should be close 
to the bottom-Yukawa one.
Another prediction relates to the decay mode $h \to Z\gamma$, 
which is enhanced only slightly, 
$\Gamma^{\rm IS} (h \to Z \gamma) = 1.07 \times \Gamma^{\rm SM} (H \to Z\gamma)$,
unlike $h \to \gamma\gamma$.
Last but not least, the LHC might potentially discover 
the top-Higgs resonance $h_t$, if lucky.
For details, see Ref.~\cite{Hashimoto:2012qe}. 

I would like to emphasize that the window of the composite Higgs models 
is still open. Stay tuned!


\bigskip 

\end{document}